\documentclass{webofc}
\usepackage[varg]{txfonts}   % Web of Conferences font
\usepackage[dvipsnames]{xcolor}
\usepackage{xcolor}
\usepackage{hyperref}
\usepackage[utf8]{inputenc}
\usepackage{amsmath}
\usepackage{amsfonts}
\usepackage{amssymb}
\usepackage{bbold}

\definecolor{nicered}{rgb}{0.7,0.1,0.1}
\definecolor{nicegreen}{rgb}{0.1,0.5,0.1}
\definecolor{red}{rgb}{1.0, 0, 0}
\definecolor{niceblue}{rgb}{0,0,0.8}
\definecolor{red}{rgb}{1.0, 0, 0}
\hypersetup{colorlinks=true,citecolor=nicegreen,linkcolor=nicered,urlcolor=nicered}

\newcommand{\beq}{\begin{equation}}
\newcommand{\eeq}{\end{equation}}

\def\vev#1{\left\langle #1\right\rangle}

\newcommand{\eq}[1]{Eq.~(\ref{#1})}
\newcommand{\eqs}[2]{Eqs.~(\ref{#1})--(\ref{#2})}
\renewcommand{\[}{\left[}
\renewcommand{\]}{\right]}
\renewcommand{\(}{\left(}
\renewcommand{\)}{\right)}
\begin{document}
%
%\preprint{DESY 19-193}

%\begin{minipage}{.45\linewidth}
    \begin{flushright}
      \text{DESY 19-193}
    \end{flushright}
%  \end{minipage}
\title{Flavour Violating Axions}

\author{
\firstname{Luca} 
\lastname{Di Luzio}\inst{1}\fnsep\thanks{Talk given at the Workshop ``Flavour changing and conserving processes'' 29--31 August '19 (Capri Island, Italy)
%.
\email{luca.diluzio@desy.de}
} 
%    \and
%        \firstname{Second author} \lastname{Second author}\inst{2}\fnsep\thanks{\email{Mail address for second
%             author if necessary}} \and
%        \firstname{Third author} \lastname{Third author}\inst{3}\fnsep\thanks{\email{Mail address for last
%             author if necessary}}
        % etc.
}

\institute{Deutsches Elektronen-Synchrotron DESY,
Notkestra{\ss}e 85, D-22607 Hamburg, Germany 
%\and
%           the second here 
%\and
%           Last address
          }

\abstract{%
I review the physics case for flavour violating axions. 
In particular, I argue that relaxing the assumption of the universality of the Peccei-Quinn 
current opens up new pathways, including:  
the relaxation of the Supernova bound on the axion mass,  
%(nucleophobia), 
a possible connection with the Standard Model flavour puzzle 
and 
the experimental opportunity of discovering the axion via flavoured axion searches. 
%and $iv)$ the exotic possibility of a heavy $\mathcal{O}(\text{MeV})$ axion. 
}
\maketitle
\tableofcontents
\clearpage 

\section{Introduction}
\label{sec:intro}

Axion physics has entered a new, experimentally driven phase. 
In the recent years we have witnessed the emergence of several new experimental proposals, 
which (in different stages of development) promise to open for exploration 
regions of parameter space considered unreachable until few years ago 
(for a recent experimental overview see Ref.~\cite{Irastorza:2018dyq}). 
It is hence timely, on the theory side, to reassess axion properties beyond standard scenarios. 
In this contribution, I first review some basics aspects of axion physics. The emphasis is put 
on the fact that axion couplings are inherently UV dependent, 
so that a wide experimental program should keep this into account, 
regardless of theoretical prejudice. 
In particular, I will focus on a commonly adopted, although not strongly motivated, assumption 
which consists in the universality of the Peccei-Quinn (PQ) current. 
I will argue, both with old examples and more recent ones, that relaxing the latter 
assumption
is motivated by various phenomenological and theoretical arguments. 
Most importantly, it offers a unique experimental handle which consists in flavour-violating (FV) 
axion transitions. 
The latter turn out to be complementary to standard axion searches 
and potentially competitive with astrophysical bounds.

\section{Axion couplings: model-independent vs.~model-dependent}
\label{sec:axionprop}

The essence of the axion solution of the strong CP problem is a 
new spin-0 field $a(x)$, the axion, endowed with a pseudo-shift symmetry $a \to a + \alpha f_a$ 
that is broken only by the operator 
\begin{equation} 
\label{eq:aGGtilde}
\frac{a}{f_a} \frac{\alpha_s}{8\pi} G \tilde G \, .
\end{equation}
While the QCD $\theta$ term can be rotated away via an appropriate 
choice of $\alpha$, a Vafa-Witten theorem \cite{Vafa:1984xg} ensures that in a 
vector-like theory such as QCD, 
$E(\vev{a} = 0) \leq E(\vev{a} \neq 0)$, 
thus solving the strong CP problem. 
The origin of the $a G \tilde G$ operator in \eq{eq:aGGtilde} can be 
traced back in the existence of a QCD-anomalous, global U(1)$_{\rm PQ}$ 
symmetry, whose spontaneous breaking delivers the axion as a  
pseudo-Nambu-Goldstone boson.  

\begin{figure}[h]
\centering
%\sidecaption
\includegraphics[width=13cm]{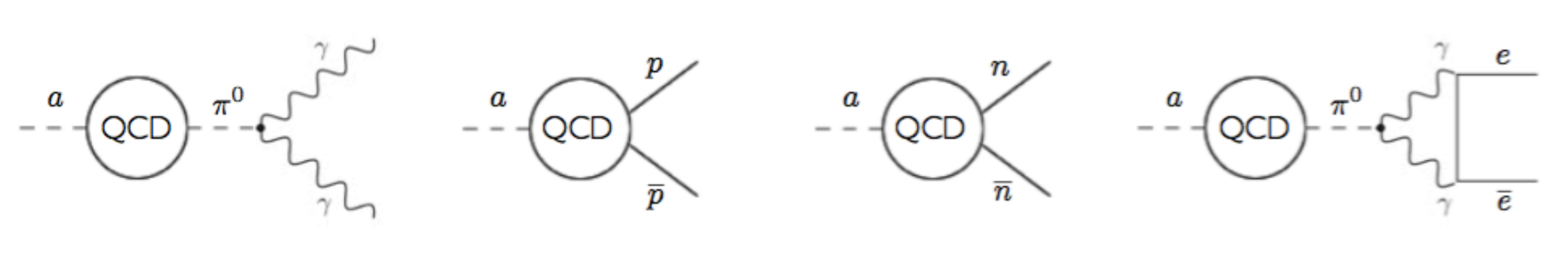}
\caption{Model-independent axion couplings. 
The blob stands for non-perturbative QCD dynamics.}
\label{fig:axioncouplMI}       
\end{figure}

The $a G \tilde G$ operator  represents the 
smoking-gun signature of the PQ mechanism 
and determines the so-called model-independent properties of the axion field, 
which are given solely in terms of the axion decay constant $f_a$. 
These are the axion mass 
$m_a = 5.691(51) \, \text{$\mu$eV} \, (10^{12} \, \text{GeV} / f_a)$ \cite{Gorghetto:2018ocs},  
and the axion couplings to 
photons, nucleons and electrons 
(here we focus on the most relevant ones for astrophysical limits), 
arising from the 
diagrams in Fig.~(\ref{fig:axioncouplMI}).

%\vspace{-0.2cm} 
%\noindent  
Given the effective axion Lagrangian 
\beq
\label{eq:LaEFT1}
\mathcal{L}_a \supset \frac{\alpha}{8\pi} \frac{C_\gamma}{f_a} a F\tilde F 
+ \frac{C_f}{2 f_a} \partial_\mu a \bar f \gamma^\mu \gamma_5 f \, , 
\quad (f = p,n,e) \, , 
\eeq
the numerical values of the coefficients $C_{\gamma,\,p,\,n,\,e}$ can be determined via 
chiral Lagrangian techniques, as well as inputs from Lattice QCD, 
and they are found to be \cite{diCortona:2015ldu,Srednicki:1985xd,Chang:1993gm}
\beq 
C_\gamma = -1.92 (4) \, , \quad 
C_p= -0.47(3) \, , \quad 
C_n= -0.02(3) \, , \quad 
C_e= - 7.8(2) \times 10^{-6} \log \(\frac{f_a}{m_e}\) \, . 
\eeq
However, 
being the description of the effective 
operator in \eq{eq:aGGtilde} valid only until energies of the order of $f_a$, 
the theory must be UV completed. Remarkably, the UV completion of the 
axion effective Lagrangian can drastically affect the low-energy properties of the axion,  
and hence the way to experimentally probe it. 

\begin{figure}[h]
\centering
%\sidecaption
\includegraphics[width=10cm]{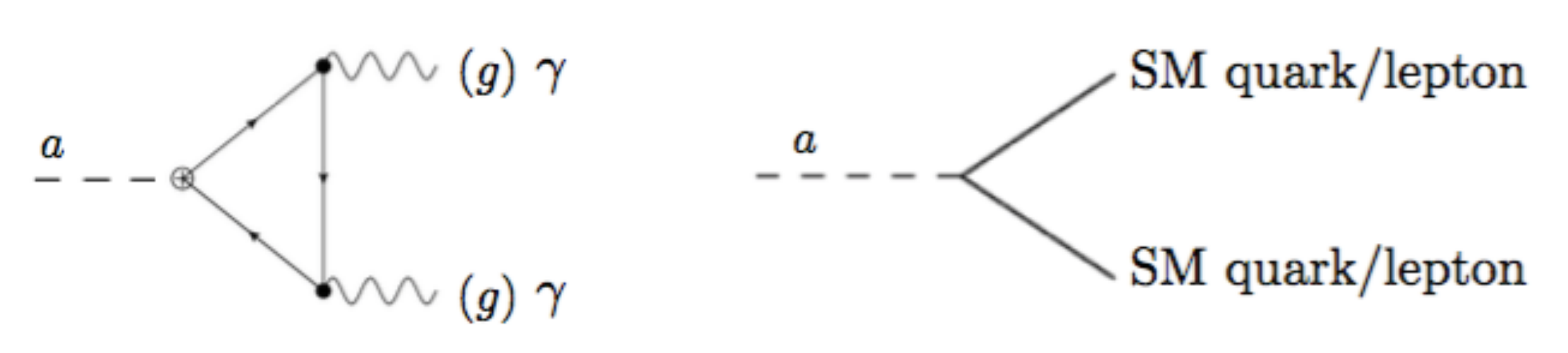}
\caption{Model-dependent axion couplings to photons and SM quarks and leptons.}
\label{fig:axioncouplMD}       
\end{figure}

There are basically two main ways in which this can happen, 
as depicted schematically in the diagrams of 
Fig.~(\ref{fig:axioncouplMD}).

%\vspace{-0.2cm} 
%\noindent
In the left diagram of Fig.~(\ref{fig:axioncouplMD}), 
the PQ-charged colored fermions responsible for generating the 
$aG\tilde G$ operator 
can also lead to a direct QED-anomalous contribution 
to $a F\tilde F$, 
if the new fermions they are charged under U(1)$_{\rm EM}$. Then the axion coupling to photons 
gets modified into $C_\gamma = E/N - 1.92 (4)$, where $E/N$ is a 
group theory factor which depends on the quantum numbers of the 
fermions running in the loop (see e.g.~Refs.~\cite{DiLuzio:2016sbl,DiLuzio:2017pfr} 
for 
phenomenologically 
motivated ranges of $E/N$). 

The other possibility, depicted in the right diagram of Fig.~(\ref{fig:axioncouplMD}), 
is that the axion interacts directly with the Standard Model (SM) fermions, which are charged under 
the U(1)$_{\rm PQ}$.  
In this case, the axion effective interaction can be written as 
(keeping for the sake of illustration only SM quarks)
\beq 
\label{eq:partialJPQ}
\frac{\partial_\mu a}{v_{\rm PQ}} J^\mu_{\rm PQ} 
= \frac{\partial_\mu a}{v_{\rm PQ}} \[ 
%v_{\rm PQ} \partial^\mu a 
- \bar Q_L \mathcal{X}_{Q_L} \gamma_\mu Q_L
- \bar u_R \mathcal{X}_{u_R} \gamma_\mu u_R
- \bar d_R \mathcal{X}_{d_L} \gamma_\mu d_R 
\] \, ,
\eeq
where $J^\mu_{\rm PQ}$ is the conserved (up to anomalies) 
PQ current, depending on the U(1)$_{\rm PQ}$ charges. 
The latter are denoted by $\mathcal{X}_{Q_L,\, u_R,\, d_R}$, 
which are diagonal (in general, non-universal) matrices.
After going to the mass basis: $u_L \to V_{u_L} u_L$, etc., and using the relation 
$f_a = v_{\rm PQ} / (2N)$ between the axion decay constant and the PQ-breaking order parameter, 
we can recast \eq{eq:partialJPQ} as 
\beq
\label{eq:partialJPQ2}
\frac{\partial_\mu a}{2 f_a} 
\bar \psi_i (C^V_{\psi_i \psi_j} + C^A_{\psi_i \psi_j}\gamma_5) \gamma^\mu \psi_j \, ,
\eeq
where mass eigenstates are denoted as $\psi_i = \{ u_i, d_i \}$ and 
we have introduced the vector\footnote{The diagonal vector couplings 
do not contribute to on-shell physical processes, as it can 
be seen upon integrating by parts and using the equations of motion. 
} 
and axial couplings
\begin{align}
\label{eq:CVAu}
C^{V,\,A}_{u_i u_j} &= \frac{1}{2N} \( V^\dag_{u_L} \mathcal{X}_{Q_L} V_{u_L}  
\pm V^\dag_{u_R} \mathcal{X}_{u_R} V_{u_R} \)_{ij} \, , \\ 
\label{eq:CVAd}
C^{V,\,A}_{d_i d_j} &= \frac{1}{2N} \( V^\dag_{d_L} \mathcal{X}_{Q_L} V_{d_L}  
\pm  V^\dag_{d_R} \mathcal{X}_{d_R} V_{d_R} \)_{ij} \, .
\end{align}
Note that the unitary flavour matrices are only subject to the constraint 
$V_{\rm CKM} = V^\dag_{u_L} V_{d_L}$. 
%, unless extra theoretical bias is invoked. 

A common assumption, 
as e.g.~in the original Peccei-Quinn-Weinberg-Wilczek (PQWW) 
\cite{Peccei:1977hh,Peccei:1977ur,Weinberg:1977ma,Wilczek:1977pj}
and the Dine-Fischler-Srednicki-Zhitnitsky (DFSZ) \cite{Zhitnitsky:1980tq,Dine:1981rt}
axion models, consists in the universality of the PQ current, 
i.e.~$\mathcal{X}_{Q_L,\, u_R,\, d_R} \propto \mathbb{I}_3$.\footnote{We remind the reader 
that instead in Kim-Shifman-Vainshtein-Zakharov (KSVZ) \cite{Kim:1979if,Shifman:1979if} type of models 
the SM fields are not charged under the U(1)$_{\rm PQ}$. 
} 
This implies flavour blind axion interactions: 
the axion interacts in the same way with SM fermions having the same gauge quantum numbers 
and, moreover, off-diagonal entries in \eq{eq:partialJPQ2} trivially vanish.

\section{A lesson from flavour}
\label{sec:lessonflavour}

The simplest axion model featuring only two Higgs doublets for the breaking of the U(1)$_{\rm PQ}$, 
also known as PQWW model, was ruled out quite soon after its conception,  
due to a combination of beam dump experiments and rare meson decays 
such as $K \to \pi a$ \cite{Hall:1981bc} 
and $\text{Quarkonia} \to \gamma a$ \cite{Wilczek:1977zn}.  
For instance, radiative decays of Quarkonia 
to $\gamma a$ normalized to leptonic modes can be written at the leading order as \cite{Wilczek:1977zn} 
\beq 
\label{eq:quarkoniaga}
\frac{\mathcal{B}(J/\psi \to \gamma a)}{\mathcal{B}(J/\psi \to \mu \mu)} = \frac{g_c^2}{2 \pi \alpha} \, , \qquad 
\frac{\mathcal{B}(\Upsilon \to \gamma a)}{\mathcal{B}(\Upsilon \to \mu \mu)} = \frac{g_b^2}{2 \pi \alpha} \, ,  
\eeq
where $g_c = m_c C^{A}_{cc} / f_a$ (in the notation of \eq{eq:partialJPQ2}) and similarly for $g_b$.
In the PQWW model one has 
\beq 
\label{eq:gcandgb} 
g_c = \frac{m_c}{v} \frac{1}{\tan\beta} \, , \qquad 
g_b = \frac{m_b}{v} \tan\beta \, ,
\eeq
with $v = 246$ GeV and $\tan\beta = \vev{H_u}/\vev{H_d}$. 
Since the couplings in \eq{eq:gcandgb} cannot be simultaneously suppressed, 
bounds from Quarkonia alone
would have been sufficient to rule out the PQWW model 
(for a historical account see Sect.~3 in Ref.~\cite{Davier:1986ps}). 

However, a main assumption behind the PQWW model consisted in the universality of the PQ current.
At the beginning of the 80's it seemed more natural to keep this assumption, 
and extend the model by adding a SM-singlet field in order to 
decouple the PQ breaking from the electroweak scale.\footnote{On the contrary, 
it can be argued that imposing $f_a \gg v$ leads to two well-known naturalness issues: 
the hierarchy problem and the PQ-quality issue.} 
Models of these type became known as \emph{invisible} axion models 
and led to the standard KSVZ/DFSZ benchmarks.  
In the same years, the GSI anomaly \cite{Schweppe:1983yv,Clemente:1983qh,Cowan:1985cn} 
(a sharp peak in the $e^+$ spectrum of heavy ion collisions) 
which could be interpreted in 
terms of an $\mathcal{O}(\text{MeV})$ axion with dominant decay mode $a \to e^+ e^-$, 
triggered a lot of interest in trying to explain that in terms of \emph{variant} axion models, 
with a non-universal axion 
mainly coupled to first generation fermions in order to 
avoid bounds from Quarkonia decays. 
It actually took almost a decade, starting from the original PQWW proposal, 
to rule out non-universal PQWW variants, 
by combining informations from rare $\pi$ and $K$ decays \cite{Bardeen:1986yb}. 
Even nowadays, there is an interesting claim 
that under certain conditions (among which 
the non-universality of the PQ current) 
an $\mathcal{O(\text{MeV})}$ axion is not obviously ruled out \cite{Alves:2017avw}.\footnote{This is not in contradiction 
with the previous statement, since the UV completion of this axion is not of the PQWW type.} 

\section{Non-universal Peccei-Quinn axion models}
\label{sec:lessonflavour}

We explore now some consequences of non-universal PQ 
axion models, focussing on some recent developments 
such as the possibility of suppressing the 
axion coupling to protons and neutrons (nucleophobia), 
which allows in turn to relax the Supernova (SN) bound on the axion mass. 
We next make some considerations on possible connections of non-universal 
PQ scenarios with the SM flavour puzzle and 
summarize the status of flavoured axion searches.

\subsection{Nucleophobia}
\label{sec:nucleophobia}

The axion coupling to nucleons (cf.~\eq{eq:LaEFT1}) 
%(protons and neutrons) 
can be computed using a 
non-relativistic effective field theory where nucleons are at rest and the axion 
is treated as an external current (see \cite{diCortona:2015ldu} for details). 
In particular, one obtains 
\begin{align} 
\label{eq:CppCn}
C_p + C_n &= (c_u + c_d) (\Delta_u + \Delta_d) - 2\delta_s \, , \\ 
\label{eq:CpmCn}
C_p - C_n &= (c_u - c_d) (\Delta_u - \Delta_d) \, , 
\end{align}
where 
$c_{u, \, d}$ are derivative axion couplings to the axial current of valence quarks in the nucleon,  
defined in the basis where the $aG\tilde G$ has been rotated away (cf.~\eq{eq:LwaGGtilde2}), 
%$c_u \equiv C^A_{u_1u_1}$ and
%$c_d \equiv C^A_{d_1d_1}$ (cf.~\eq{eq:partialJPQ2}),  
$\delta_s \approx 5\%$ encodes effects from sea quarks in the nucleon, 
and $\Delta_{u, \, d}$ are nucleon matrix elements 
which are extracted from experiments 
and/or the Lattice. Hence, if we want to simultaneously suppress 
both the axion-nucleon couplings, say at the level of $10\%$, we need to impose 
$c_u + c_d = 0$ and $c_u - c_d = 0$, regardless of the specific 
value of $\Delta_{u,d}$. 

It is instructive to see why achieving such cancellation 
is not possible in standard KSVZ and DFSZ models. 
Let us consider the axion Lagrangian in \eq{eq:partialJPQ}, 
restricted to first generation up and down quarks, and ignoring for the moment flavour mixing. 
Including also the $aG\tilde G$ 
term, the latter reads 
\beq
\label{eq:LwaGGtilde1}
\mathcal{L}_a \supset \frac{a}{f_a} \frac{\alpha_a}{8 \pi} G \tilde G 
+ \frac{\partial_\mu a}{v_{\rm PQ}} \[ \mathcal{X}_u \bar u \gamma^\mu \gamma_5 u 
+ \mathcal{X}_d \bar d \gamma^\mu \gamma_5 d \] \, ,
\eeq
where 
$2\mathcal{X}_u \equiv \mathcal{X}_{Q_1} - \mathcal{X}_{u_1}$ and 
$2\mathcal{X}_d \equiv \mathcal{X}_{Q_1} - \mathcal{X}_{d_1}$
are the PQ charges expressed in terms of the chiral ones (for notational simplicity, 
we have suppressed the chirality indices in the right-hand sides). 
The $aG\tilde G$ term can be rotated away via a field-dependent 
2-flavour quark transformation: $q \to \exp{(i\gamma_5 Q_a a(x) / (2f_a))} \, q$, 
with $q=(u,d)^T$ and $Q_a = \text{diag}(m_d / (m_u + m_d), m_u / (m_u + m_d))$.
The latter choice ensures that the axion has no mass mixing with the neutral pion \cite{Georgi:1986df}. 
In the new basis, \eq{eq:LwaGGtilde1} reads 
\beq
\label{eq:LwaGGtilde2}
\mathcal{L}_a \supset 
\frac{\partial_\mu a}{2 f_a} \Bigg[ 
\underbrace{\( \frac{\mathcal{X}_u}{N} - \frac{m_d}{m_u + m_d} \)}_{c_u} 
\bar u \gamma^\mu \gamma_5 u 
+ \underbrace{\( \frac{\mathcal{X}_d}{N} - \frac{m_u}{m_u + m_d} \)}_{c_u} 
\bar d \gamma^\mu \gamma_5 d \Bigg] \, .
\eeq
Hence, the nucleophobic conditions can be recast as 
\begin{align} 
\label{eq:cppcn}
0 = c_u + c_d &= \frac{\mathcal{X}_u + \mathcal{X}_d}{N} - 1 \, , \\ 
\label{eq:cpmcn}
0 = c_p - c_n &= \frac{\mathcal{X}_u - \mathcal{X}_d}{N} - \frac{m_d-m_u}{m_u + m_d}  \, . 
\end{align}
While the second condition can always be implemented via a tuning (see below), 
the real bottleneck is the first one, since $\mathcal{X}_u = \mathcal{X}_d = 0$ 
in KSVZ models,  
while in DFSZ models one has 
$N = \frac{1}{2} n_f (2 \mathcal{X}_{Q_1} - \mathcal{X}_{u_1} - \mathcal{X}_{d_1}) 
= n_f (\mathcal{X}_u + \mathcal{X}_d)$, with $n_f = 3$ denoting the number of families.  

The above no-go theorem for standard axion models suggests itself a 
possible way out \cite{DiLuzio:2017ogq}: 
if the total anomaly factor were equal to that of the first family,  
i.e.~$N = N_1 = \mathcal{X}_u + \mathcal{X}_d$, the first condition 
would be automatically satisfied. 
To simplify the discussion, let us assume a $2+1$ structure such that 
the PQ charges of the first two generations the same. 
Then the condition we would like to impose, $N\equiv N_1+N_2+N_3=N_1$, 
simply implies $N_1 = N_2 = -N_3$. It is remarkably simple to obtain the latter 
in terms of a renormalizable Yukawa Lagrangian featuring two Higgs doublets 
$H_{1,2} \sim (1,2,-1/2)$. 
For instance, 
\beq 
\label{eq:LYukPQ}
\mathcal{L}_Y = \bar Q_3 u_3 H_1 + \bar Q_3 d_3 \tilde H_2 + \ldots 
+ \bar Q_2 u_2 H_2 + \bar Q_2 d_2 \tilde H_1 + \ldots 
+ \bar Q_1 u_1 H_2 + \bar Q_1 d_1 \tilde H_1 + \ldots \, ,
\eeq
where we have suppressed Yukawa couplings and the dots stand for 
off-diagonal operators which are necessary in order to obtain a realistic CKM mixing, 
compatibly the overall PQ charge assignments.\footnote{This include as well a 
scalar potential communicating the PQ breaking to the two Higgs doublets via a SM 
singlet scalar, in a fashion similar to universal DFSZ models. In the absence 
of texture zeros and for a 2+1 structure for the PQ charges, all the possible Yukawa structures 
leading to nucleophobia have been 
classified in Ref.~\cite{DiLuzio:2017ogq}.  
The opposite case of maximal numbers of texture zeros 
has been discussed instead in Ref.~\cite{Bjorkeroth:2018ipq}.} 
Denoting by $\mathcal{X}_{1,\,2}$ the PQ charges of $H_{1,\,2}$, 
from \eq{eq:LYukPQ} we have 
$N_3 = \frac{1}{2} (2 \mathcal{X}_{Q_3} - \mathcal{X}_{u_3} - \mathcal{X}_{d_3}) 
= \frac{1}{2} (\mathcal{X}_1 - \mathcal{X}_2)$ and 
$N_2 = \frac{1}{2} (2 \mathcal{X}_{Q_2} - \mathcal{X}_{u_2} - \mathcal{X}_{d_2}) 
= \frac{1}{2} (\mathcal{X}_2 - \mathcal{X}_1) = - N_3$, having swapped the role of 
$H_1$ and $H_2$ for the third and second/first generations. 
Hence, $N = N_1 = \frac{1}{2} (2 \mathcal{X}_{Q_1} - \mathcal{X}_{u_1} - \mathcal{X}_{d_1}) 
= \mathcal{X}_u + \mathcal{X}_d$ and \eq{eq:cppcn} is automatically satisfied 
in terms of charge assignments. 

Coming to the second nucleophobic condition in \eq{eq:cpmcn}, this can be expressed 
as a condition on $\tan\beta = \vev{H_2} / \vev{H_1}$. In order to see that, 
let us first impose the orthogonality between the PQ 
and hypercharge currents: $Y(H_1) \mathcal{X}_1 v_1^2 + Y(H_2) \mathcal{X}_2 v_2^2 = 0$, 
and hence $\mathcal{X}_1 / \mathcal{X}_2 = - \tan^2\beta$, implying no kinetic 
mixing between the axion and the $Z$ boson.  
Using $\mathcal{X}_{u} - \mathcal{X}_{d} 
= \frac{1}{2} (\mathcal{X}_{d_1} - \mathcal{X}_{u_1}) =  \frac{1}{2} (\mathcal{X}_{2} + \mathcal{X}_{1})$
%(where the first step comes from the definitions below \eq{eq:LwaGGtilde1}, 
%while the second step from \eq{eq:LYukPQ}) 
and $N = N_1 = \frac{1}{2} (2 \mathcal{X}_{Q_1} - \mathcal{X}_{u_1} - \mathcal{X}_{d_1}) 
= \frac{1}{2} (\mathcal{X}_{2} - \mathcal{X}_{1})$, we can recast \eq{eq:cpmcn} 
as $(\mathcal{X}_{2} + \mathcal{X}_{1}) / (\mathcal{X}_{2} - \mathcal{X}_{1}) = \cos^2\beta - \sin^2\beta = (m_d - m_u) / (m_u + m_d) \approx 1/3$, 
which is satisfied for $\tan\beta \approx 1/\sqrt{2}$. 

To sum up, with a single 
$\mathcal{O}(10\%)$ tuning,\footnote{The level of tuning useful for nucleophobia is limited by the remainder 
$2\delta_s \approx 10\%$ in \eq{eq:CppCn}, which sets an irreducible contribution 
to nucleon couplings. In principle, flavour mixing 
(neglected in the present simplified discussion) 
also enters diagonal axion couplings when going to the mass basis (cf.~\eqs{eq:CVAu}{eq:CVAd}), 
and it can be used to further cancel the $\delta_s$ contribution \cite{DiLuzio:2017ogq}.} 
$\tan\beta \approx 1/\sqrt{2}$, 
one can simultaneously suppress both the axion coupling to neutrons and protons, 
which allows in turn to relax the SN bound by one order of magnitude compared to standard 
KSVZ/DFSZ scenarios, e.g.~in KSVZ models the often quoted bound reads 
$m_a \lesssim 0.02$ eV \cite{Tanabashi:2018oca}. 
However, recent analyses of the axion emissivity from the SN core 
hint to a weakening of the bound by a factor of a few \cite{Chang:2018rso,Carenza:2019pxu}. 

It should be noted that in the 
construction above the axion-electron 
coupling, $g_{ae}$, turns out to be generically sizeable,\footnote{Having a sizeable $g_{ae}$ 
with a somewhat relaxed SN bound improves the fit of the cooling anomalies \cite{Giannotti:2017hny}.} 
and additional strong bounds on 
$m_a$ are obtained from white-dwarf cooling rates and red giants evolution in globular clusters. 
Hence, in order to effectively relax astrophysical bounds on $m_a$, also $g_{ae}$ must be somewhat 
suppressed. This can be done at the cost of an extra tuning with flavour mixing in the leptonic 
sector \cite{DiLuzio:2017ogq}, or, more elegantly, by introducing a leptonic Higgs doublet $H_3$ 
which allows for a correlated cancellation of the axion coupling to nucleons and electrons 
via a single tuning \cite{Bjorkeroth:2019jtx}. 
This kind of scenarios have been dubbed \emph{astrophobic}, in the sense that 
all the main astrophysical constraints can be relaxed up to one order of magnitude, 
allowing for axions as heavy as $m_a \sim 0.2$ eV. 
On the other hand, the axion coupling to photons remains generically sizeable, 
e.g.~$E/N = \{ -4/3, 2/3, 8/3, 14/3 \}$ in the models of Ref.~\cite{DiLuzio:2017ogq},
so that the next generation of helioscopes, such as IAXO \cite{Armengaud:2019uso}, 
will be able to probe their parameter space.

\subsection{Connections with the Standard Model flavour puzzle} 
\label{sec:SMflavourp}

We have seen in the previous section that nucleophobia provides 
a clear case for the non-universality of the PQ current, 
but it does not fix the size of flavour mixing 
entering into the axion couplings 
to SM fermions when going to the mass basis (cf.~\eqs{eq:CVAu}{eq:CVAd}). 
The diagonalizing flavour matrices, subject only to the constraint $V_{\rm CKM} = V^\dag_{u_L} V_{d_L}$, 
are in principle free parameters
which can be measured in flavour violating (FV) processes involving the axion. 
Extra theoretical bias, possibly in connection with the SM flavour puzzle, can help 
in providing an organizing principle for the flavour structure. 
A particularly natural realization is to identify the U(1)$_{\rm PQ}$ with the 
horizontal U(1) symmetry responsible for the Yukawa hierarchies 
\cite{Davidson:1981zd,Wilczek:1982rv,Davidson:1983fy,Davidson:1984ik}.\footnote{Gauging the horizontal flavour group can also 
lead to an accidental global U(1) whose spontaneous breaking delivers a flavoured  
arion \cite{Berezhiani:1983hm} or a flavoured axion \cite{Berezhiani:1985in,Berezhiani:1989fs,
Berezhiani:1989fp}. 
} 
Models of Froggatt-Nielsen \cite{Froggatt:1978nt} type have recently 
regained some attention (see e.g.~\cite{Ema:2016ops,Calibbi:2016hwq}), 
and they typically predict axion flavour transitions controlled by the CKM matrix, 
%i.e.~$C^{A,V}_{ij} \sim (V_{\rm CKM})_{ij}$, 
although subject to irreducible 
$\mathcal{O}(1)$ uncertainties common also to other flavour models. 
A different kind of approach \cite{Bjorkeroth:2018ipq,Bjoorkeroth:2019ndr}, 
based on requiring the maximal number of textures zeros 
(compatibly with SM fermion masses and mixings)
that can be enforced via a family dependent U(1)$_{\rm PQ}$ at the renormalizable level, 
allows instead to completely fix\footnote{Up to the usual dependence from 
$f_a$ and $\tan\beta$, as in universal DFSZ models.} 
the axion couplings to SM fermions, including off-diagonal ones, 
in terms of a fit to SM fermion masses and mixings.

\subsection{Flavoured axion searches} 
\label{sec:FAsearches}

Rare FV decays with invisible and light final states allow to 
probe the off-diagonal axion couplings $C^{A,V}_{\psi_i \psi_j}$ in \eq{eq:partialJPQ2}. 
A collection of bounds can be found for instance in Refs.~\cite{Feng:1997tn,Bjorkeroth:2018dzu,Ziegler:2019gjr}. 
The strongest limits on FV axion couplings to quarks come from
$K^+ \to \pi^+ a$.
Comparing the theoretical prediction with the current limit from E949/E787~\cite{Adler:2008zza} gives
$m_a < 2 \cdot 10^{-5} \ \text{eV} / \, |C^V_{sd}|$,  
which for maximal mixing (i.e.~$C^V_{sd} = \mathcal{O}(1)$) is about three orders of magnitude stronger 
than typical astrophysical bounds. 
Hence, $K^+ \to \pi^+ a$ clearly provides a golden channel for FV axion 
searches.\footnote{It should be noted that pseudo-scalar meson decays such as $K \to \pi a$
are sensitive only to the vectorial part of the quark current, 
since $\langle \pi | \bar s \gamma_\mu \gamma_5 d | K \rangle = 0$ 
by the Wigner-Eckart theorem. In order to set bounds on $C^A_{sd}$  
one has to resort to other FV processes, as for example $K^0-\bar K^0$ mixing, 
which however are much weaker compared to the limits on the 
vectorial counterpart from meson decays. 
In principle, this would leave open a possibility 
in order to evade the strong constraints from 
$K \to \pi a$ in the presence of large mixing, 
if one could cook up a model with $C^V_{sd} \ll C^A_{sd}$.   
%Similar considerations apply to $B$ mesons as well. 
}  
NA62 is expected to improve the limit on $\mathcal{B}(K^+ \to \pi^+ a)$ by a factor of 
$\sim$70~\cite{Anelli:2005ju,Fantechi:2014hqa}, thus 
strengthening the axion mass bound 
by a factor $\sim$8.  
The next most sensitive process in the quark sector is 
$B^+ \to K^+ a$, whose present limit from CLEO~\cite{Ammar:2001gi} 
corresponds to 
$m_a < 9 \cdot 10^{-2} \ \text{eV} / \, |C^V_{bs}|$, 
close to astrophysical limits for maximal mixing. 
This latter bound could be presumably strengthened by a factor $\sim$10 at BELLE II \cite{Abe:2010gxa}.  
A similar, but slightly weaker bound, 
is obtained for $B^+ \to \pi^+ a$, which involves the coupling $C^V_{bd}$. 
On the other hand, bounds on processes involving other quark transitions 
are about three orders of magnitude smaller 
(see Table 2 in Ref.~\cite{Bjorkeroth:2018dzu}), 
and hence not competitive with astrophysical limits, even for maximal mixing. 
In the charged lepton sector, 
the strongest limits come from (30 years old) searches for 
$\mu\to e\gamma a$~\cite{Bolton:1988af,Goldman:1987hy}. 
They yield $m_a < 3 \cdot 10^{-3} \ \text{eV} / \, (|(C^V_{\mu e})^2 + |(C^A_{\mu e})^2|)^{1/2}$, 
competitive with astrophysical bounds in the case of sizeable flavour mixing. 
This bound is likely to be improved by one order of magnitude at the MEG \cite{Renga:2018fpd} 
and Mu3e \cite{Blondel:2013ia} experiments at PSI. 
On the other hand, bounds on $\tau$-$\mu$ and $\tau$-e transitions from ARGUS 
\cite{Albrecht:1995ht} yield axion mass limits which are still well-below astrophysical ones.

\section{Conclusions}
\label{sec:concl}

Relaxing the assumption of the universality of the PQ current opens up the parameters 
space of DFSZ models, by introducing 34 new real parameters only in the quark sector 
(most of which are related to the diagonalizing U(3) matrices in \eqs{eq:CVAu}{eq:CVAd}). 
These new parameters arise, in some sense, within the SM (since they come from the 
diagonalization of the Yukawas that we write in the SM Lagrangian), but the SM is ``blind'' to them.  
By extending the SM with the axion at low-energy, it is somewhat artificial to require that flavour mixing 
beyond the CKM is unphysical and we should rather let experiments to decide.  
In the meanwhile, it is worth to speculate about the possible consequences of the non-universality of the 
PQ current, as I have done in the present contribution. 

To conclude, I cannot resist from making an unfortunate analogy: ``DFSZ$\,=\,$cMSSM''. 
The DFSZ model corresponds, in some sense, to the constrained version 
of the MSSM with universal soft terms. 
There is however an important difference, 
while large flavour violating effects in low-energy SUSY were often considered as a curse 
(flavour problem), in the case of the axion they rather come as a blessing (flavour opportunity).   

\section*{Acknowledgments}

I thank the organizers of the FCCP19 workshop, and in particular Giancarlo D'Ambrosio, 
for the warm hospitality. 
I also thank Fred Bj\"{o}rkeroth, Federico Mescia, Enrico Nardi, Paolo Panci and Robert Ziegler 
for an enjoyable collaboration on flavoured axions. 
This work is supported by the Marie Sk\l{}odowska-Curie Individual Fellowship 
grant AXIONRUSH (GA 840791).

\bibliography{bibFCCP2019.bib}

\end{document}